\documentstyle[11pt,pasp,twoside,psfig]{article}
\def\edcomment#1{\iffalse\marginpar{\raggedright\sl#1\/}\else\relax\fi}
\reversemarginpar

\begin{document}

\title{X-Ray Observations on the Galactic Center Region}
\author{Katsuji Koyama}

\affil{Department of Physics, Graduate School of Science,
Kyoto University}

\begin{abstract}
This paper reports  on the early   Chandra view of the Galactic center (GC) activities.
The massive black hole Sgr~A$^*$  is extremely faint, while more bright diffuse X-ray 
emission is prevailing  in the  circumnuclear  disk.
Another high temperature plasma  is  
found in the Sgr A East shell.¡¡ This may indicate that Sgr A East is a supernova remnant, 
although  no clear X-ray shell is found.
A hint of non-thermal X-ray filaments is found, suggesting the presence
of an acceleration site of extremely high-energy cosmic rays.
The giant molecular cloud Sgr B2 is established to be an X-ray 
reflection nebula, possibly arising from the past Sgr A activities.
Chandra further discovered  
high temperature shells, suggesting multiple supernova explosions near  the GC
region.
\end{abstract}

\section{Introduction}

The Galactic center (GC)  and its vicinity  exhibit highly complex  features,  possibly 
originated  from  the activity of either a putative massive black  hole (MBH) or  violent star 
bursts (e.g. Genzel et al. 1994).  Since  X-rays usually accompany these activities,  
observations in this band, 
particularly in the hard X-rays, are vital.
With the Ginga satellite, Koyama et al.(1989) discovered  strong iron line emission from an
extended region near the  GC with  a thermal energy of 10$^{54}$ ergs
and  proposed that either an  energetic explosion occurred at the MBH (Sgr~A$^*$), or 
multiple supernova explosion took  place  within the past 10$^5$ years. 

The ASCA observations confirmed the Ginga results and furthermore  found 
more detailed structures: diffuse thin-thermal X-rays   
of a few arcmin region in  the Sgr A complex 
with the peak at Sgr~A$^*$ and 
largely extended emission ($\sim 1 \deg$)  with lower surface brightness.    
The X-ray spectrum in  a $\sim 1\deg \times 1 \deg $  region
exhibits many emission lines, 
particularly triplet-line structure is found at 6.9, 6.7 and 6.4  keV, which are 
attributable to K-shell lines from  H, He-like and neutral irons.  
The former two are due to high temperature plasmas, while the latter neutral line 
indicates the presence of cold gas in a strong X-ray field. However no  bright 
X-ray source at Sgr~A$^*$ and/or its vicinity  has been found. 

The "X-ray quiet" Sgr~A$^*$ is in  sharp contrast to the
surrounding high temperature plasmas.
The 6.4 keV line emission gives a hint to connect the quiet Sgr~A$^*$ 
and the active environment.  
This emission  is found to be clumpy; the brightest regions are  the Sgr B2 cloud
and the Galactic Center Radio Arc (Koyama et al. 1996).  
The Sgr C and Clump II radio sources are
also found to emit 6.4 keV lines (Murakami et al. 2001a, Sakano et al. 2000). 
The X-ray spectra and morphologies  are well explained by 
an "X-ray Reflection Nebula (XRN)" model; the X-rays are due to reflection,
photoelectric absorption and fluorescence from iron atoms. 
Since no adequately bright X-ray source was
found in the vicinity of the XRNs to fully account for the
diffuse X-ray flux,  Koyama et al. (1996) and Murakami et al.(2000) 
proposed 
a  scenario, despite its
considerable distances from the XRNs, that  Sgr~A$^*$
exhibited an X-ray outburst by a possible surge of accretion on the MBH in the near past, 
and is currently in a quiescent accretion phase.

For further X-ray study,  high resolution imaging
spectroscopy on the GC region is essentially important.
The Chandra ACIS-I  
is an ideal instrument with
a spatial resolution of 0.5$\arcsec$ (the size of the CCD pixel) 
and a  reasonably large  field of view of 17$'\times$ 17$'$.
This paper reports the ACIS-I results
of  key regions  near the GC  based on published papers
by Baganoff et al. (2001), Maeda et al. (2001) and Murakami et al.
(2001), and our new analysis of the archive data.
\section{Sgr A$^*$ and Its Close Vicinity}
\begin{figure}[t]
\hspace*{1cm}
\begin{tabular}{cc}
\psfig{file=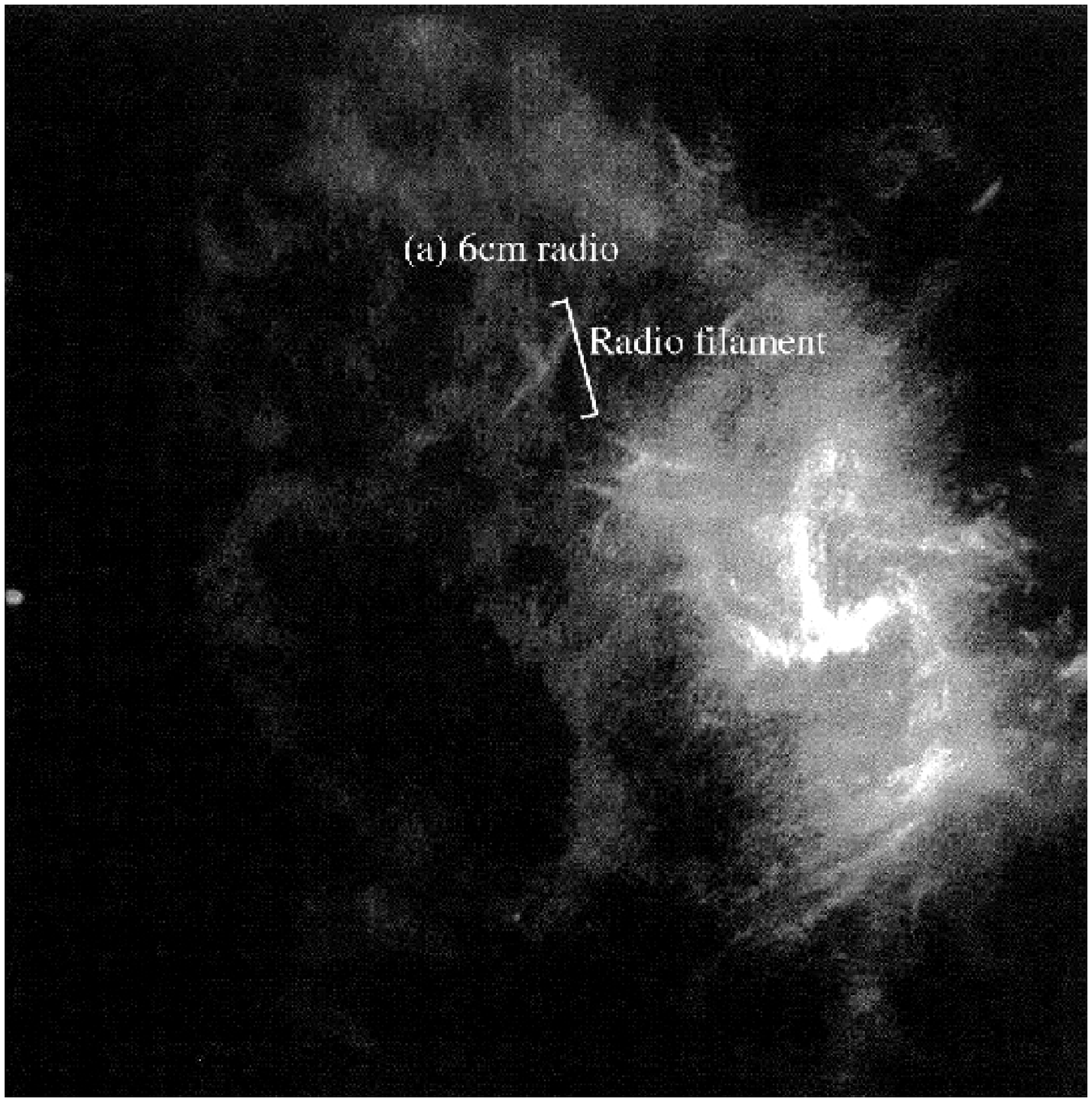,width=0.42\textwidth,clip=} &
\psfig{file=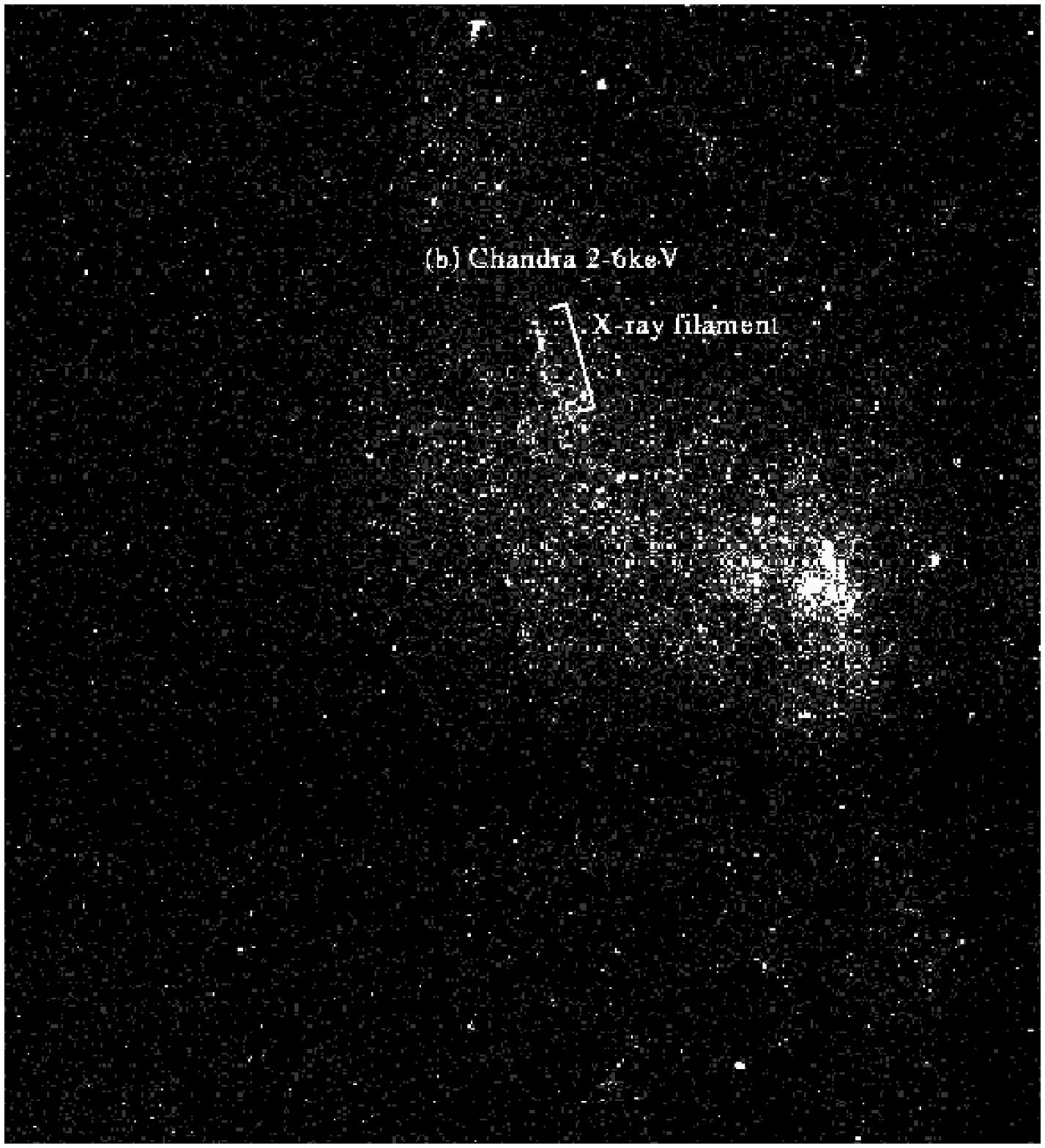,width=0.42\textwidth,clip=}
\end{tabular}
\caption{The radio and X-ray images near Sgr A$^*$ of a $\sim 3.5' \times 3.5'$ field.
North is up and East is left.
Left: The 6cm radio continuum image taken from Genzel et al. 1994 (figure 2.2), 
which was made from the data provided by Lo.
Right: The X-ray image in the 2--6~keV band obtained with Chandra
(archive: Obs ID=242).
An X-ray filament is discovered near a radio filamentary structure.}
\end{figure}

Figure 1 (left) is the radio map of the GC region of $\sim 3.5'\times 3.5'$. 
The large arc structure in the upper region
is  a part of the Sgr A East shell.  The smaller shell near the right of the image 
is called the "Circumnuclear Disk" (CND), and inside is the "Mini-Spiral" with the center 
of Sgr~A$^*$.
Figure 1 (right) is the Chandra ACIS image  produced from the archive data (Obs ID=242).
In X-rays, the brightest point source in this region is Sgr~A$^*$ (near the lower right corner).
Baganoff et al.(2001) reported that  Sgr~A$^*$  has  a power-law spectrum of 
$\Gamma=\sim$2.7 with an X-ray luminosity of $\sim2\times 10^{33}$ergs s$^{-1}$,  
two orders of magnitude lower 
than the upper limit measured with the previous 
instruments.  Some   point-like sources have IR 
counterparts,  but  others are unknown sources.
An even brighter  source  is diffuse emission in the CND.  
The X-ray spectrum is a thin thermal plasma of $\sim$1.3 keV 
temperature with a luminosity of 
$\sim 2\times 10^{34}$ergs s$^{-1}$, 
ten times brighter than Sgr~A$^*$ (Baganoff et al.2001).
%
\begin{figure}
\hspace*{2cm}
\psfig{file=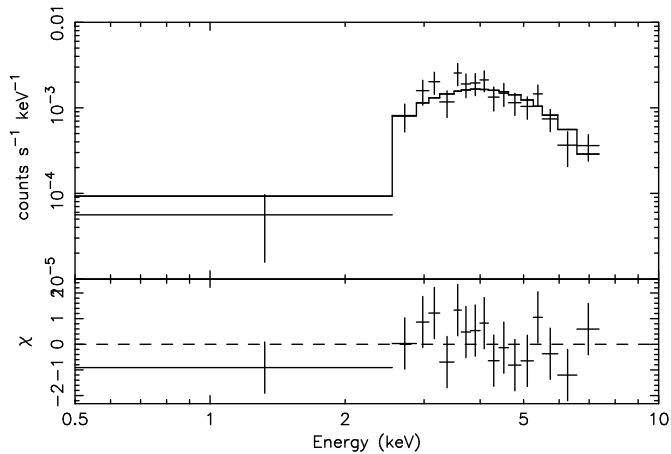,width=0.66\textwidth,clip=}
\caption{The ACIS spectrum of a filament in the Sgr A East SNR (see figure 1).}
\end{figure}

\section{Sgr A East }

As is shown in figure 1 (right), diffuse X-ray emission is found
inside the Sgr A East radio shell. 
The X-ray spectrum is  fitted  with a 2-keV plasma model of over-abundant metals 
(Maeda et al. 2001). The presence of the metal-rich, high temperature plasma inside the 
non-thermal radio shell 
indicates that Sgr A  East is a young  SNR.  
No soft X-ray emission is detected from the radio shell,
possibly due to the heavy  absorption to this direction. 
For  simplicity, we assume a Sedov model for Sgr A East with 
canonical  explosion energy of $10^{51}$ ergs. Then  the reduced age of the SNR
is  $\sim 10^3$ years.  A notable conclusion  is that the ambient 
density is as high as $10^{2\sim3}$ cm$^{-3}$.  This is consistent with
the fact that the radio spectrum of the Sgr A complex shows a turnover at 90 cm, 
which means that dense  ionized gas is  prevailing around the Sgr A complex.  
The X-ray absorptions of the GC sources 
are all $N_{\rm H} \sim 10^{23}$ H cm$^{-2}$ (Baganoff et al. 2001).
On the other hand,  optical extinction of IR stars is $A_{\rm v} \sim$ 30 mag, or 
$\sim 6\times 10^{22}$  ($N_{\rm H} =1.8\times 10^{21} A_{\rm v}$).
This apparent difference of $N_{\rm H}$  between the optical and X-ray estimations
can  be explained by the  presence of the  dense ionized gas region with a dimension
of  about 10 pc (Maeda et al. 2001).

We found a filamentary structure in the northern region of this
young SNR.  The X-ray spectrum shows no emission line, and can be 
fitted with
a power-law model of photon index $\Gamma=\sim$2.5 (figure 2).
In the radio image of Sgr A  East (figure 1 left), we find a possible radio 
counterpart near the X-ray filament.
Another X-ray filamentary  structure  is found near 
the Galactic Center Radio Arc in the other archive data of the GC region 
(archive: Obs ID=945).  
These facts suggests that the X-ray 
filaments  may be due to  synchrotron 
radiation, as is seen in some  shells of young  SNRs, like SN1006 (Koyama et al. 1995).  
Since the synchrotron photon energy is given by:
$E_{\rm ph}$ = $2{\rm [keV]}(B/1{\rm mG})(E_{\rm e}/10{\rm TeV})^2$, 
detection of non-thermal X-rays is good evidence that  extremely high-energy
particles are accelerating near the GC.    
The energy-loss time-scale is:
$T_{\rm syn} = (B/1{\rm mG})^{-2} (E_{\rm e}/10{\rm TeV})^{-1}$ [yr], 
therefore the filaments should be very young 
possibly $10\sim100$ years.   
Protons, of which the energy-loss is negligible, may be accelerated to even higher energy, 
if the  magnetic field of the filaments is as high as  m-Gauss.
In may be noted that the Japanese Air shower group, AGASA found an enhancement 
of cosmic rays to the GC  direction at very high energies of $\geq 10^{18}$eV 
(Hayashida et al. 1999).

\section{New X-Ray SNRs}

\begin{figure}
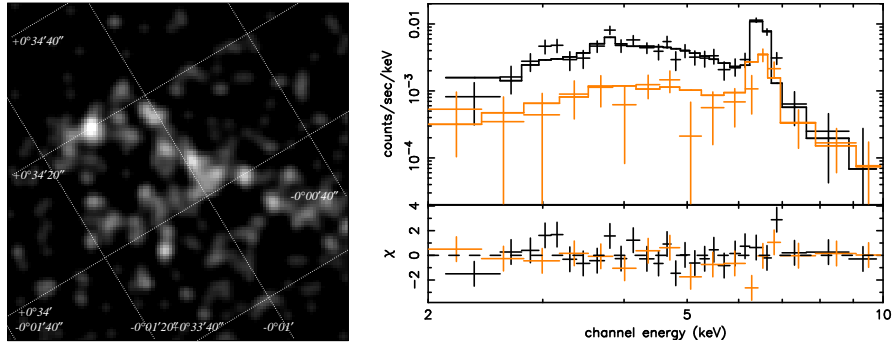
 
\hspace*{1cm}
\begin{tabular}{cc}
\psfig{file=koyama_katsuji_fig3a.epsi,width=0.34\textwidth} &
\psfig{file=koyama_katsuji_fig3b.epsi,width=0.51\textwidth}
\end{tabular}
\caption {Left: The Chandra image of a new SNR in the 6.0-7.0 keV band. 
Right: The X-ray spectra of ASCA (gray line) and Chandra (black line).}
\end{figure}

The superior spatial resolution of Chandra revealed that  the diffuse 
X-rays are  clumpier than it was  observed  with  ASCA. In order to investigate
the origin of the clumpy emission, we have analyzed some of the X-ray clumps.  
At the west of  the Sgr B2
cloud, we found a small ring of $\sim 10\arcsec$-diameter (figure 3 left, 
Senda et al. 2001). 
The X-ray spectrum (figure 3 right) is well fitted with 
a high-temperature (6 keV) thin-thermal plasma in non-ionization equilibrium, hence the source
would be a young SNR. A massive progenitor star would make 
a high-density gas ring in the equatorial plane  and  would be shock-heated
by the supernova ejecta, which is essentially the same model
as SN1987A (Burrows et al. 2000). In fact, 
if we placed this source in the Large Magellanic Cloud, we would  observe 
an X-ray ring of $\sim 2.5\arcsec$ radius,  5 times larger than  SN1987A.  
From the X-ray luminosity ($\sim 10^{34}$ erg s$^{-1}$), the ring volume (0.6 pc-radius with
0.2 pc of thickness), and the best-fit ionization parameter (log $n\tau \sim $ 10.3), 
we can estimate the ionization
age to be $\tau \sim$  2$\times 10^9$ sec.
Thus the X-ray ring may  be the youngest SNR in our Galaxy, heated-up less than 100 years ago. 

Another diffuse structure with a strong iron line is found at the south of Sgr A East 
associated with a 
non-thermal radio ``wisp'' (Sgr A-E) (Ho et al. 1985).  
The X-ray spectrum is fitted with a
thin  thermal plasma.  We see a faint X-ray excess in a $\sim$2$'$-radius circle along  the 
X-ray shell, suggesting this to be  another  new young SNR.  We thus expect that
many new SNRs will be discovered  
in the GC region in deep Chandra observations, 
which will  account for a significant fraction of the diffuse X-ray emission.
\section {The Role of Sgr A East to the X-ray Reflection Nebula}

\begin{figure}
\begin{tabular}{cc}
\psfig{file=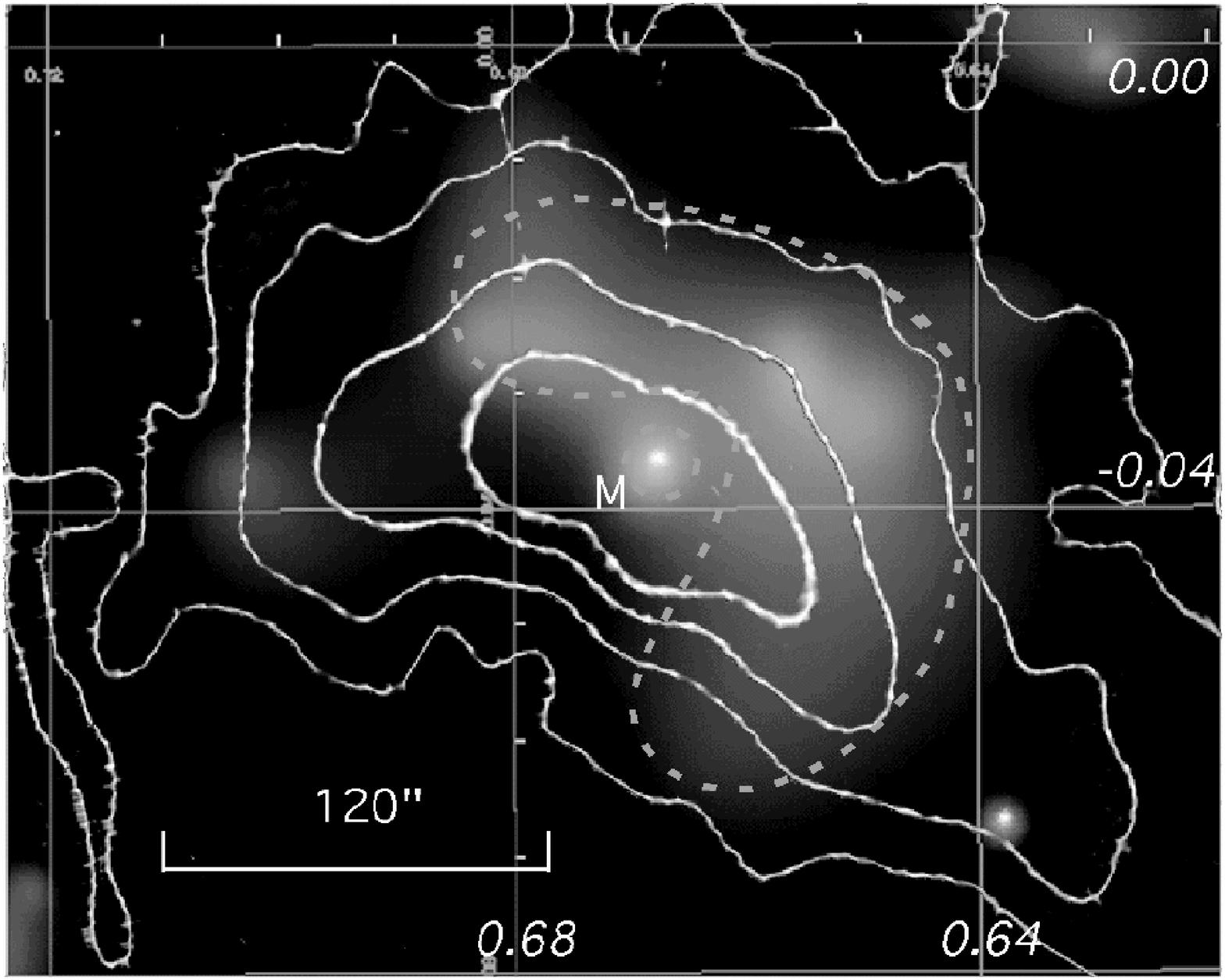,width=0.45\textwidth,clip=} &
\psfig{file=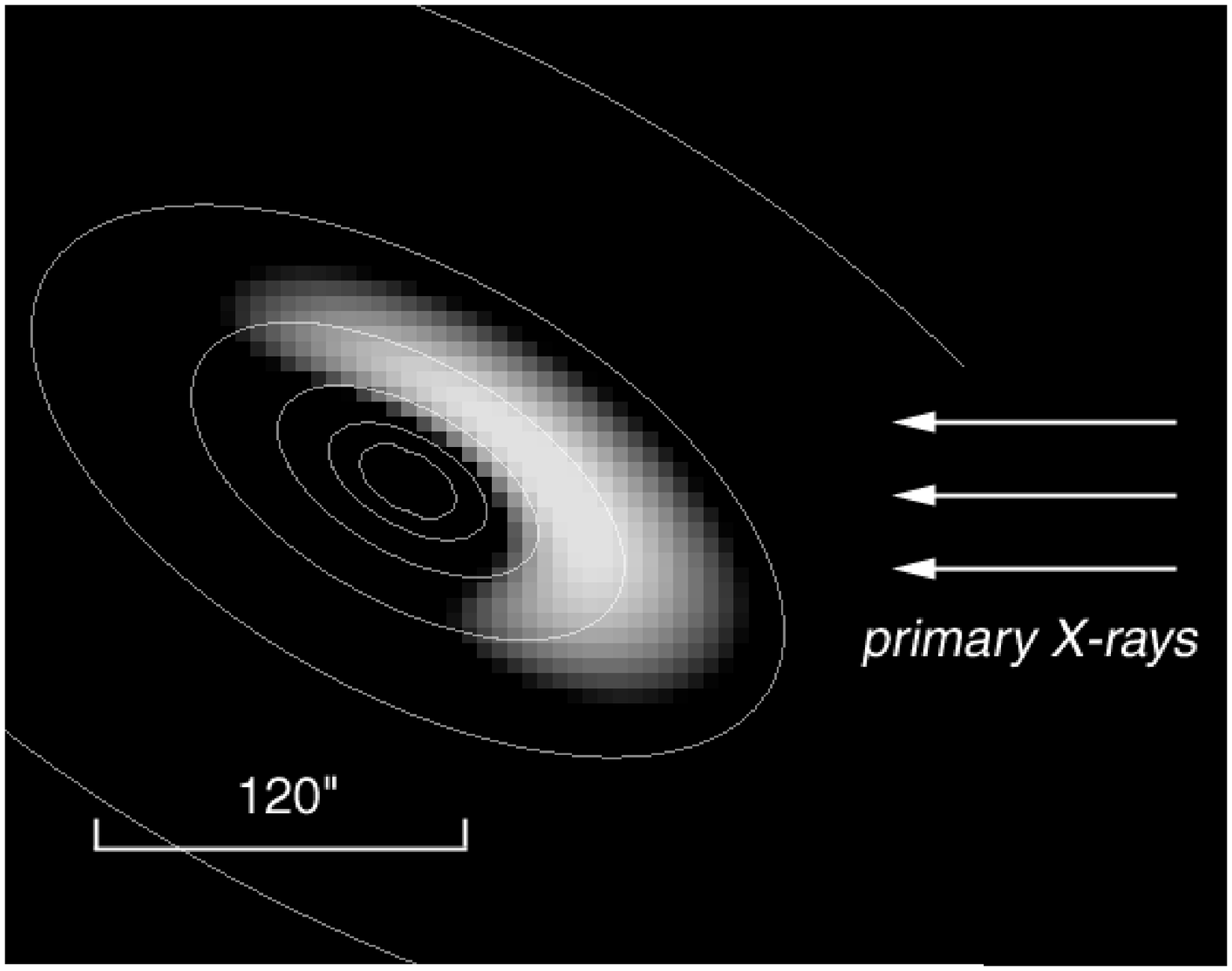,width=0.45\textwidth,clip=}
\end{tabular}
\caption{Left: The Chandra ACIS-I image around the
Sgr~B2 cloud in the 6.0--7.0 keV band. The contour shows the density
distribution of the molecular cloud (Sato et al. 2000).
Right: Simulated fluorescent line image based on an XRN model with the primary X-ray source at 
the direction to the GC (right). The contours show the density distribution. }
\end{figure}

 Among the new X-ray SNRs, Sgr A East may be  particularly important.
The shock wave of Sgr A East in the dense medium made a dense shell with a speed 
of $\sim 10^{3}$ km sec$^{-1}$.   
Since the projected distance 
between Sgr~A$^*$ and the radio shell is about 1pc, the encounter would have been 10$^{3}$  
years ago.  
What happened when the dense shell passed across the MBH (Sgr~A$^*$) ?
Maeda et al. (2001) interpreted that 
the MBH may had been  activated via  Bondi-Hoyle accretion from the dense gas shell
to emit strong X-rays for a few 100 years. 
The strong radiation field may have ionized 
the surrounding dense gas, which is still surviving as is noted in section 3.

At the same time, the bright X-rays from the MBH started a long distance travel. 
After $\sim$300 years, the strong X-rays  arrived  at a giant molecular cloud Sgr B2, 
then produced 
a fluorescent iron line (the 6.4 keV line)  with a concave shape pointing at the GC as is 
given in figure 4 (left). 
A simulated image based on the XRN model
given in figure 4 (right) is in  good agreement with the observed image.   
The X-ray spectrum (figure 5) shows strong K$_\alpha$ and K$_\beta$ 
lines with exactly the proper flux ratio of the laboratory atomic 
data.  A deep absorption edge at 7.1 keV is also present. These results  are fully 
consistent  with  the fluorescent and reflection  scenario.
With this scenario, we may expect many other XRNs near the GC. In fact ASCA
found several other candidates, which are the Galactic Center Radio Arc (Koyama et al. 1996), 
Sgr C (Murakami et al. 2001a) and  Clump II (Sakano et al. 2000).  Chandra observations
on these objects are highly required.\\
\\
The author expresses his sincere thanks to F. Baganoff, Y. Maeda,  
H. Murakami, J. Yokogawa and A. Senda for their helps and useful information
in preparing this paper.
%
\begin{figure}
\hspace*{2cm}
\psfig{file=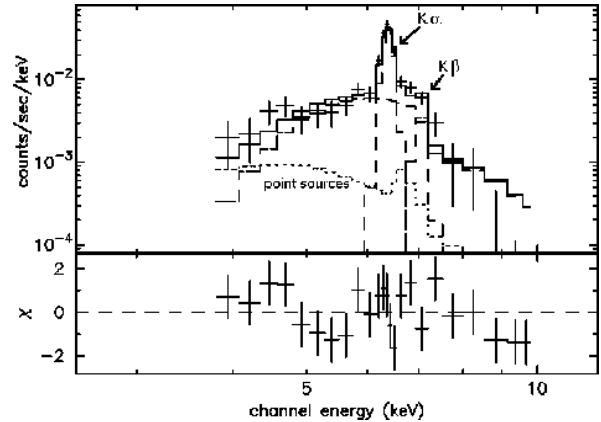,width=0.65\textwidth}
\caption{The ACIS spectrum of the diffuse X-rays from the
Sgr~B2 cloud. The dashed line indicates K$_\alpha$ and K$_\beta$
lines, while the dotted
line shows the contribution of the point sources to the diffuse
spectrum.}
\end{figure}
\section{References}
\begin{quote}
Baganoff, F., et al. 2001, \apj, in press (astro-ph 0102151)\\
Burrows, D. N.,  et al. 2000, \apj, 543L, 149 \\
Genzel, R.,Hollenbach, D.,\& Townes, C. H. 1994, Rep. Prog. Phys, 57, 417 \\ 
Hayashida, N., et al. 1999, Astroparticle Physics, 303, 311\\
Ho, P.T.P., Jackson, J.M., Barrett, A.H. \& Armstrong, J.T. 1985, \apj, 288, 575 \\
Koyama, K., et al. 1989, Nature, 339, 603\\
Koyama, K., et al. 1995, Nature, 378,255\\
Koyama, K., et al. 1996, \pasj, 48, 249 \\
Maeda, Y., et al. 2001, \apj, in press (astro-ph/0102183)\\
Murakami, H., et al. 2000, \apj, 534, 283\\
Murakami, H., et al. 2001a, \apj, 550, 297\\
Murakami, H., Maeda, Y., \& Koyama, K. 2001b, \apj, in press (astro-ph/0105273)\\
Sakano, M. 2000, phD thesis, Kyoto University\\
Sato, F., Hasegawa, T., Whiteoak, J.B., \& Miyawaki, R. 2000, \apj, 535, 857\\
Senda, A., Murakami, H., \& Koyama, K. 2001, \apj, submitted \\
\end{quote}
\end{document}